\documentclass[5p, times]{elsarticle}
\usepackage{amssymb}
\usepackage{amsmath}
\usepackage{amsfonts}
\usepackage{graphicx}
\usepackage{xcolor}
\usepackage{xspace}
\usepackage{ulem}
\usepackage{mathtools}
\usepackage{hhline}

\usepackage[T1]{fontenc}
\usepackage{newtxtext}
\usepackage[varg]{newtxmath}
\biboptions{sort&compress}

\begin{document}

\title{Warm Little Inflaton becomes Dark Energy}

\author[av]{Jo\~{a}o G.~Rosa} \ead{joao.rosa@ua.pt} 
\author[av]{Lu\'{\i}s B. Ventura} \ead{lbventura@ua.pt}

\address[av]{Departamento de F\'{\i}sica da Universidade de Aveiro and CIDMA,  Campus de Santiago, 3810-183 Aveiro, Portugal}
 
\date{\today}

\begin{abstract}
We present a model where the inflaton field behaves like quintessence at late times, generating the present phase of accelerated expansion. This is achieved within the framework of warm inflation, in particular the Warm Little Inflaton scenario, where the underlying symmetries guarantee a successful inflationary period in a warm regime sustained by dissipative effects without significant backreaction on the scalar potential. This yields a smooth transition into a radiation-dominated epoch, at which point dissipative effects naturally shut down as the temperature drops below the mass of the fermions directly coupled to the inflaton. The post-inflationary dynamics is then analogous to a thawing quintessence scenario, with no kination phase at the end of inflation. Observational signatures of this scenario include the modified consistency relation between the tensor-to-scalar ratio and tensor spectral index typical of warm inflation models,  the variation of the dark energy equation of state at low redshifts characteristic of thawing quintessence scenarios, and correlated dark energy isocurvature perturbations.
\end{abstract}

\maketitle


Since the discovery of cosmic acceleration by the observations of Type Ia supernovae (SNe) \cite{Riess:1998cb,Perlmutter:1998np}, dark energy (DE) has become an essential part of the standard cosmological model. Whether DE is a complement to the well-known energy budget of the Universe  described by General Relativity or generated by modifications of the latter remains to be seen. 

Microscopic/first principles descriptions of either scenario have remained elusive, but these appear crucial if one is to have a more complete understanding of the problem of cosmic acceleration. We think, however, that this is not emphasized enough: in the particular case of quintessence, a scalar field description of dynamical dark energy, one does not usually motivate neither the origin nor the characteristics of such a homogeneous scalar field. 

Interestingly, it is also hypothesized that a homogeneous scalar field, the inflaton $\phi$, generates a phase of accelerated expansion in the early Universe, thereby solving the homogeneity and horizon problems and providing the seeds for structure formation and Cosmic Microwave Background (CMB) anisotropies \cite{Baumann:2009ds}. One could then ask: can the same field generate these two similar stages of accelerated expansion, happening at such different energy scales?

Inflaton-quintessence unification was originally suggested in \cite{Spokoiny:1993kt,Peebles:1998qn} and investigated in more general settings \cite{Dimopoulos:2000md, Huey:2001ae, Majumdar:2001mm, Ng:2001hs, Dimopoulos:2001ix, Nunes:2002wz, Sami:2004xk, Rosenfeld:2005mt, Zhai:2005ub, Cardenas:2006py, BuenoSanchez:2006fhh, Membiela:2006rj, Rosenfeld:2006hs, Neupane:2007mu, Bento:2008yx, BasteroGil:2009eb, Bento:2009zz, Dias:2010rg, Hossain:2014xha, Hossain:2014zma, Geng:2015fla, Khalifeh:2015tla, Haro:2015ljc, deHaro:2016ftq, Guendelman:2016kwj, deHaro:2017nui, Rubio:2017gty, Geng:2017mic, Ahmad:2017itq, vandeBruck:2017voa, Dimopoulos:2017tud, Casas:2018fum, Wetterich:2019qzx, Akrami:2017cir}. It may have noteworthy cosmological consequences as the generation of a baryon asymmetry \cite{DeFelice:2002ir,Bettoni:2018utf} and the modification of the constants of nature \cite{Carroll:1998zi}.


These models rely on reheating mechanisms, as gravitational particle production \cite{Ford:1986sy}, which occurs due to the changing background geometry (sudden transition between inflation and kination) or instant preheating \cite{Felder:1999pv}, which occurs through a two-stage mechanism $\phi \rightarrow \chi \rightarrow \psi$, where $\chi$ is either a boson or fermion of mass $m_{\chi} = g \phi$ (with $ 10^{-7} < g < 1 $) and $\psi$ are light fermions. It was pointed out in \cite{Peebles:1998qn,Felder:1999pv} that gravitational particle production is very inefficient. Furthermore, it requires a large number of beyond the Standard Model scalar fields to dilute the contribution of gravitons to the number of relativistic degrees of freedom during Big Bang Nucleosynthesis (BBN) \cite{Peebles:1998qn} and may lead to other cosmological problems, as large isocurvature fluctuations or overproduction of dangerous relics \cite{Felder:1999pv}. Instant preheating appears a better suited solution to the "graceful exit" problem but its near-future detection appears unlikely.

In this work, we present a model where the inflaton accounts for the current DE density within the warm inflation paradigm \cite{Berera:1995wh, Berera:1995ie}. This shares various similarities with the scenario that we have proposed in \cite{Rosa:2018iff} where the inflaton accounted for the all the dark matter in the Universe, within the Warm Little Inflaton model \cite{Bastero-Gil:2016qru}, which constitutes the simplest successful realization of warm inflation in quantum field theory. This scenario exhibits three main features: (i) the inflaton is generated by the collective spontaneous breaking of a U(1) symmetry \cite{ArkaniHamed:2001nc}, which, in combination with a discrete interchange symmetry, removes all corrections to the inflaton effective potential (both at zero and finite temperature), (ii) the inflaton interacts with other fields only during inflation, becoming stable and sterile afterwards and (iii) these interactions sustain a non-negligible (albeit sub-dominant) radiation bath during inflation through dissipative effects, leading to a smooth transition into a radiation-dominated era with no need for a reheating process \cite{Berera:2008ar}. Here we show that these same features can naturally lead to quintessential inflation.

Given the similarities of this model with that of \cite{Rosa:2018iff}, here we will only briefly discuss its main technical features and refer the interested reader to \cite{Rosa:2018iff, Bastero-Gil:2016qru} for more details. The inflaton is generated when two complex scalar fields, $\phi_1$ and $\phi_2$, with identical charges under a U(1) gauge symmetry, develop an identical non-zero vacuum expectation value, leading to collective spontaneous symmetry breaking parametrically close to the grand unification scale. While the overall phase of the fields is the Goldstone boson that becomes the longitudinal polarization of the massive U(1) vector field, the relative phase between the two complex fields, the only remaining light field is the inflaton $\phi$. The vacuum manifold can be written, in the unitary gauge, as:
\begin{equation} \label{vacuum}
\phi_1= {M\over \sqrt{2}}e^{i\phi/M}~, \quad \phi_2= {M\over \sqrt{2}}e^{-i\phi/M}~.
\end{equation}
The complex scalars also interact with chiral fermion fields $\psi_{1 L,R}$ and $\psi_{2L,R}$, and we impose a discrete interchange symmetry under which $\phi_1\leftrightarrow \phi_2$ and $\psi_{1L,R}\leftrightarrow \psi_{2L,R}$, as well as a sequestering of the "1" an "2" sectors (e.g.~via additional global symmetries or localization in extra-dimensions). This yields a Lagrangian density:
\begin{eqnarray}
-\mathcal{L}_{\phi \psi} &=&{g\over\sqrt{2}}\phi_1\bar\psi_{1L}\psi_{1R}+{g\over\sqrt{2}}\phi_2\bar\psi_{2L}\psi_{2R}+\mathrm{H.c.}\nonumber\\
&=& gM \overline{\psi}_1e^{i \gamma_5 \phi/M}\psi_1 + gM \overline{\psi}_2 e^{-i \gamma_5 \phi/M}\psi_2 \: \:, \label{mWLI Lagrangian}
\end{eqnarray}
where in the last line we used the vacuum parametrization of Eq.~(\ref{vacuum}) and $\psi_{1,2}$ denote the resulting Dirac fermions, of mass $gM$ (at zero temperature). These fermions then remain light provided that the temperature during inflation is $gM \lesssim T \lesssim M$. The fact that the physical mass of the fermions is independent of the inflaton field ensures that no quantum and thermal corrections can spoil the flatness of the inflaton potential, which holds both during and after inflation. This corresponds to a modification of the original Warm Little Inflaton scenario \cite{Bastero-Gil:2016qru}, where only the leading thermal corrections to the scalar potential were absent as a result of the discrete interchange symmetry.

Note that the interchange symmetry corresponds to a $\mathbb{Z}_2$ reflection of the inflaton, $\phi\leftrightarrow -\phi$, composed with an interchange $\psi_{1} \leftrightarrow \psi_{2}$, which prevents the inflaton from decaying into any field besides the $\psi_{1,2}$ fermions. When $T < gM$, these mediators become non-relativistic and decouple from the dynamics, such that the inflaton becomes stable at late times.

During inflation, however,  $T > gM$, and the fermions can be produced through non-equilibrium dissipative effects sourced by the slow motion of the inflaton field. These arise from non-local corrections to the inflaton's effective action, which are non-zero at finite temperature despite the absence of local corrections to the scalar potential. Physically, such dissipative effects are a consequence of the inflaton's interaction with the fermions in the thermal bath, as characteristic of any dynamical system interacting with the surrounding environment.

We assume that the fermions $\psi_{1,2}$ are kept close to a local thermal equilibrium distribution, $\Gamma_\psi>H$, through Yukawa interactions with an additional light scalar field $\sigma$ and light fermion field $\psi_{\sigma}$ in the thermal bath, of the form $h \sigma \overline{\psi}_{1,2} \psi_{\sigma} + \text{H.c.}$\footnote{We do not consider direct couplings between the inflaton and the light fields since their renormalized values can be chosen to be arbitrarily small at the high energies relevant during inflation. After inflation, the fermions $\psi_{1,2}$ decouple from the effective field theory and hence do not regenerate such couplings through quantum corrections.}. Dissipative effects then lead to an additional friction term $\Upsilon\dot\phi$ in the inflaton's equation of motion, where the dissipation coefficient, $\Upsilon$, can be computed from first principles using quantum field theory at finite temperature\footnote{More details are available at https://journals.aps.org/prl/supplemental/\\10.1103/PhysRevLett.122.161301/Sup$\_$Mat$\_$Letter.pdf}
 \cite{Rosa:2018iff},
\begin{equation}
\begin{aligned}
&\Upsilon \approx  \frac{2 g^2 h^2}{(2\pi)^3} T \left(\frac{1}{5} - \ln \left( \frac{m_{\psi}}{T}\right)\right)\: \:, \: \:  m_{\psi}/T  \ll 1 ~,
\\
&\Upsilon \approx  \frac{2 g^2 h^2}{(2\pi)^3} T \sqrt{\frac{\pi}{2}}  \sqrt{\frac{m_{\psi}}{T}} e^{-\frac{m_{\psi}}{T}} \: \:, \: \:  m_{\psi}/T  \gg 1 ~. \label{LowTDC}
\end{aligned}
\end{equation}
where $m_{\psi} = \sqrt{g^2 M^2 + h^2T^2/8}$ is the thermally corrected $\psi_{1,2}$ mass. We thus have $\Upsilon\propto T$ in the high-temperature regime relevant during inflation. The warm inflation dynamics is then described by the coupled differential equations for the inflaton and radiation energy density:
\begin{align}
\ddot\phi + (3H+\Upsilon)\dot\phi + V'(\phi) = 0 ~,\nonumber\\
\dot\rho_R + 4H \rho_R = \Upsilon \dot\phi^2~. \label{Equations of Motion EFV}
\end{align}
The Hubble parameter is determined by the Friedmann equation $H^2 = (\rho_{\phi} + \rho_R)/(3 M_{\text{P}}^2)$, where $M_{\text{P}}^2 = 1/(8\pi G)$ is the reduced Planck mass, and the inflaton and radiation energy densities are given by $\rho_{\phi} \equiv \dot{\phi}^2 /2 + V(\phi)$ and $\rho_R = (\pi^2 /30) g_* T^4$, with $g_* $ denoting the number of relativistic degrees of freedom (DOF). Note that, if $\Upsilon \rightarrow 0$, i.e.~if interactions between $\phi$ and $\psi_{1,2}$ become negligible, whatever radiation density $\rho_R$ exists in the early Universe is diluted away by the inflationary expansion as in cold inflation models. This would then require an extra reheating mechanism at the end of inflation to recover the Hot Big Bang cosmology.

To completely determine the dynamics, one has to choose the inflaton potential $V(\phi)$. In the case where this is given by the simplest renormalizable potential \cite{Rosa:2018iff}, one obtains inflaton cold dark matter. To obtain inflaton dark energy, one has to choose a potential with no minima at finite field values that is flat enough at late times. As we will see, the required flatness of the potential depends on the initial conditions for the quintessence behavior, which are set at the end of the inflationary period. As a proof of concept, we adopt the same potential as \cite{Peebles:1998qn}:
\begin{equation}
V(\phi) = \left\{
\begin{array}{c} \lambda (\phi^4 + m^4) \\ \frac{\lambda m^4}{1+ \left(\frac{\phi}{m}\right)^{\alpha}} \\
\end{array} \right. ~ \begin{array}{l} \phi \leq 0 \\ \phi > 0 \end{array}\, ~,
\end{equation}
where $\lambda \sim 10^{-15}$ is fixed by the CMB anisotropies power spectrum \cite{Akrami:2018odb}, $\alpha \geq 0$ is a tunable parameter that determines the steepness of the quintessence potential and $m$ is a mass scale, fixed by present dark energy density. For a dimensionless Hubble parameter $h \approx 0.68$ and $\Omega_{\Lambda} \approx 0.69$, consistent with \cite{Aghanim:2018eyx}, the example in Fig.~1, where $\alpha = 1$, has $m \approx 3 \text{ MeV}$. Although the $\mathbb{Z}_2$ symmetry is broken in the second branch of the potential (and in fact in any quintessence model with a finite field value), the effective mass of the inflaton field is $\lesssim H_0 \sim 10^{-33} \text{ eV}$, and its interactions are mediated by heavy fields, $\psi_{1,2}$, so that $\phi$ becomes effectively stable and sterile at late times.

For large negative values of $\phi$, the quartic term dominates and slow-roll inflation occurs. For large positive values of $\phi$, the potential becomes an inverse power law, $V(\phi) \approx \lambda m^{4+\alpha}/\phi^{\alpha}$. Provided that $\phi$ is large enough after inflation, the inflaton will eventually reenter a slow-roll regime, causing a second stage of accelerated expansion.


\begin{figure}
	\centering\includegraphics[scale=0.4]{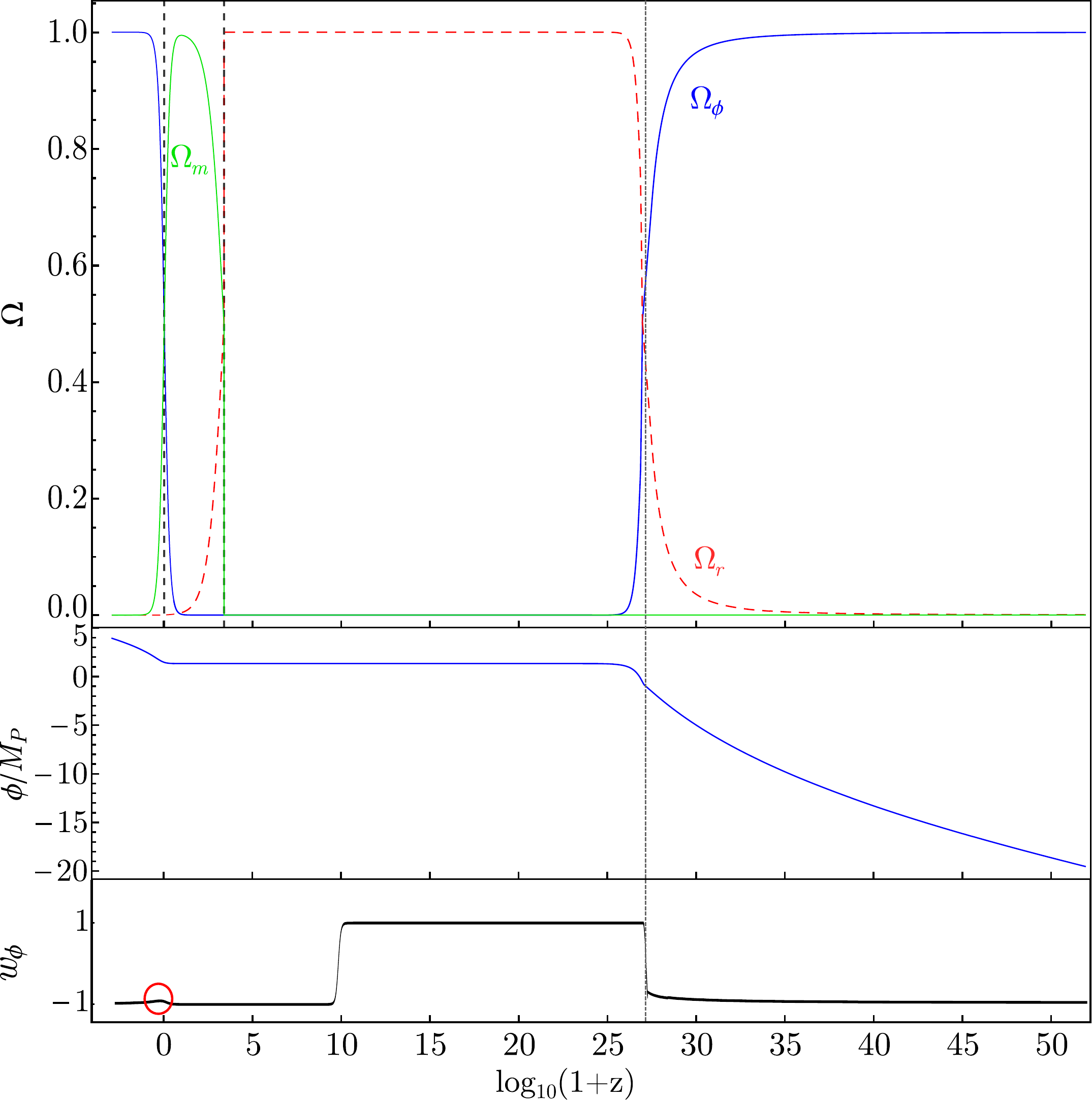} 
	\caption{{\it Top panel}: Cosmological evolution of the relative abundances  $\Omega_i$ for inflaton-quintessence (blue), radiation (red) and non-relativistic matter (green), as a function of the redshift $z=(a-1)/a$. {\it Middle and bottom panels}: Evolution of the value of the field and of the inflaton equation of state with redshift. The red circle around $z \approx 0$ signals the variation of $w_{\phi}$ at late times as detailed in Fig.~3. In this example we have chosen parameters $Q_* = 0.04,~\phi_* = -19.5M_{\text{P}}$ ($h=1.8,~g=0.35,~M \simeq 7.2 \times 10^{14} \text{ GeV}$) where $*$ denotes quantities evaluated at horizon-crossing for the chosen pivot CMB scale, so $\phi_\infty \approx 1.3 M_{\text{P}}$ and $\alpha = 1$.}
\end{figure}

A representative example of the cosmological evolution is illustrated in Fig.~1 from right to left. Since initially $K(\phi) \ll V(\phi)$, the slowly rolling inflaton field acts as an effective cosmological constant for at least $\sim 60$ e-folds, at the same time sustaining a non-negligible radiation density through the aforementioned dissipative effects, as displayed in Fig.~2: while $H$, $\Upsilon$ and $\phi$ are slowly varying, there is a non-zero steady-state solution for $\rho_R \approx \Upsilon \dot\phi^2/4H$.
Dissipative effects become strong ($Q \equiv \Upsilon/(3H) \gtrsim 1$) at the end of inflation, allowing radiation to take over as the dominant component of the energy density of the Universe at a temperature $T_R\sim 10^{13}-10^{14}$ GeV, since
\begin{equation}
	\frac{\rho_R}{V(\phi)} \simeq \frac{1}{2}\frac{\epsilon_{\phi}}{1+Q}\frac{Q}{1+Q}~,
\end{equation}
where we used the steady-state solution above and the slow-roll equations $ 3H(1 +Q)\dot\phi = - V'(\phi)$, $H^2 \approx V(\phi)/(3 M_{\text{P}}^2)$, with $\epsilon_{\phi} \equiv M_{\text{P}}^2 (V'/V)^2/2$. During inflation, $\epsilon_\phi \ll 1+Q$ and radiation is sub-dominant, while at the end of the slow-roll regime $\epsilon_{\phi} \sim 1+Q$ and radiation can smoothly take over if $Q\gtrsim 1$ \cite{Berera:2008ar}. This is in contrast with \cite{Peebles:1998qn,DeFelice:2002ir,Bettoni:2018utf} where inflation was followed by a period of kination. Concomitantly, the $\dot\rho_R$ term in \eqref{Equations of Motion EFV} becomes significant and makes the temperature drop below the mass of the $\psi_{1,2}$ fermions. Therefore, these decouple from the dynamics, making the inflaton effectively stable. Around the same time, $T/H$ becomes large enough to excite all the Standard Model DOF \cite{KolbTurner}, resulting in a temperature drop by a factor $\sim$2 and further suppressing $Q$. This is represented by the first dashed line (from right to left) in the top panel of Fig.~1. 

\begin{figure}
	\centering\includegraphics[scale=0.60]{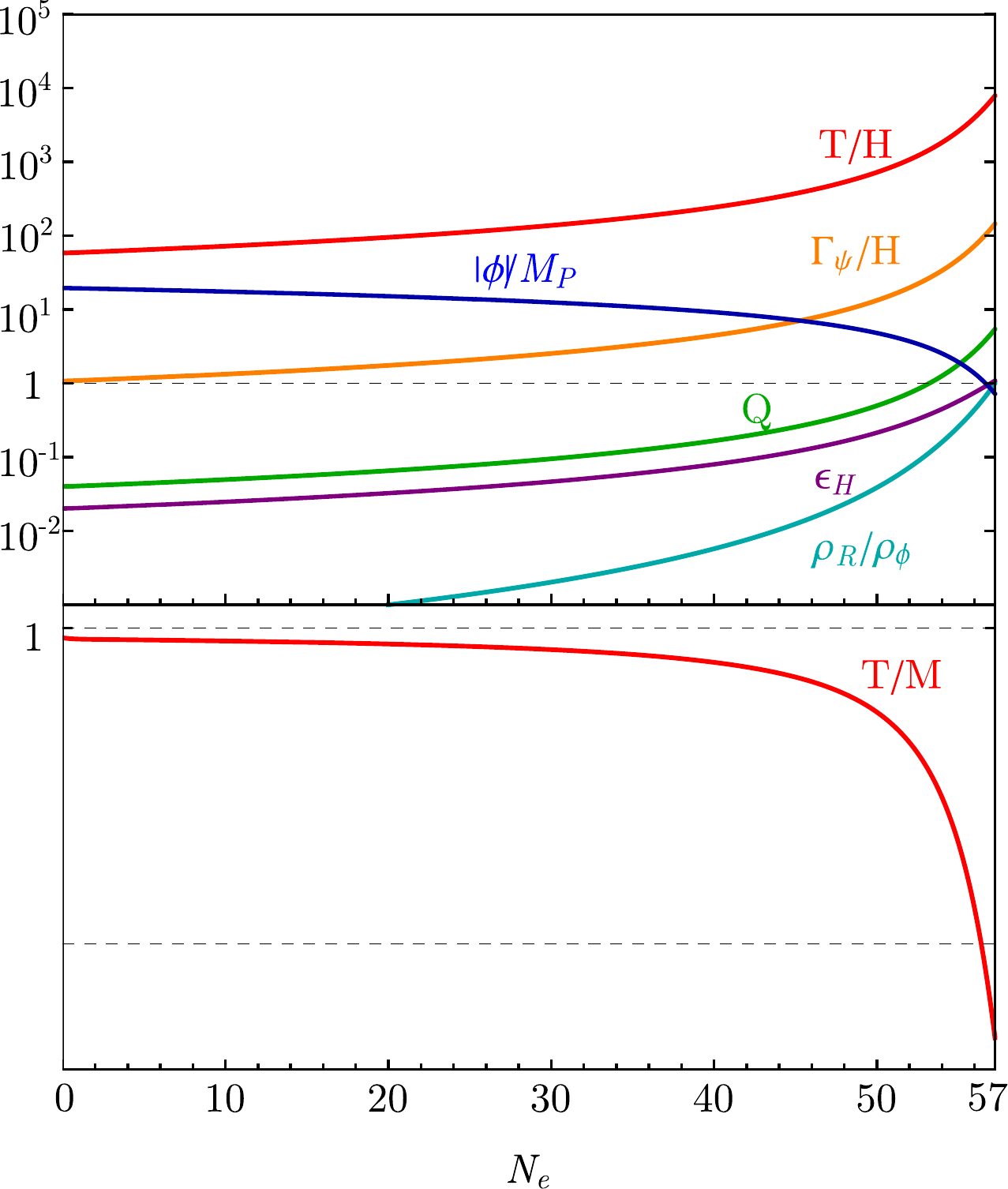} 
	\caption{ Inflationary dynamics for $V(\phi)=\lambda \phi^4$ as a function of the number of e-folds for the same choice of parameters as in Fig.~1. The total number of e-folds of inflation is $N_{\text{eq}} \approx 57$ yielding $(n_s,r)=(0.963,~2\times 10^{-3})$. The effective "reheating" temperature (at inflaton-radiation equality) is $T_R \approx 2.3 \times 10^{14} \text{ GeV}$. }
\end{figure}

Since the dissipation coefficient $\Upsilon$, given in Eq.~\eqref{LowTDC}, becomes exponentially suppressed for $m_{\psi} > T$, the inflaton field will then quickly roll from $|\phi_\text{eq}| \lesssim M_{\text{P}}$ at inflaton-radiation equality towards $\phi=0$ in an approximately quartic potential (check first dashed line from right to left in the middle panel of Fig.~1). A precise description of this rolling phase requires a numerical solution of \eqref{Equations of Motion EFV} (see, however, \cite{Greene:1997fu}\footnote{The parametric shape of $\phi(t)$ can be obtained by considering the inflaton field rolling down the quartic potential with negligible dissipation in a radiation-dominated universe. This is similar to the solution obtained in \cite{Greene:1997fu}, where $\rho_\phi$ is dominant but its equation of state is close to that of relativistic matter.}), but the important point is to determine the velocity at which the field crosses the origin, $\phi'_0\equiv \left.\dot\phi/H\right|_{\phi=0}$,  since this determines how far it can move in the second branch of the potential, as we will see below.

An instantaneous conversion of the inflaton's potential energy at the end of inflation into kinetic energy at the origin would then yield $\phi'_0\simeq \sqrt{3} M_{\text{P}}$. However, we find that the finite duration of this rolling period, where both Hubble and dissipative friction play a non-negligible role, yields somewhat smaller values. For instance, in the example shown in Fig.~1, we find $\phi'_0\simeq 1.2 M_{\text{P}}$. More generally, we find numerically that this quantity exhibits a mild dependence on the value of the dissipative ratio $Q_*$ when the relevant CMB scales become super-horizon approximately 60 e-folds before the end of inflation, yielding $\phi'_0\simeq (0.5-1.5)M_{\text{P}}$ for $Q_* \sim 10^{-3}-10^{-1}$. 

Although here we will not discuss in detail the spectrum of primordial density fluctuations in the Warm Little Inflaton scenario, referring instead the reader to the detailed analyses in \cite{Bastero-Gil:2016qru, Bastero-Gil:2017wwl, Bastero-Gil:2018uep} and references therein, we note that for a quartic potential consistency with CMB observations requires weak dissipation at horizon-crossing, in the above mentioned range. This ensures that the growth of thermal inflaton fluctuations during inflation is mild enough to yield a sufficiently red-tilted power spectrum, at the same time giving the necessary suppression of the tensor-to-scalar ratio through the thermal enhancement of scalar inflaton fluctuations when $T\gtrsim H$ at horizon-crossing. We emphasize that $Q$ is dynamical, with $Q\gtrsim 1$ at the end of inflation even if $Q_*\ll1$, as illustrated in Fig.~2.

Once the field rolls beyond $\phi > 0$, its potential energy becomes negligible and the equation of motion becomes:
\begin{equation}
	\phi'' + \left(3 + \frac{H'}{H} \right)\phi'  = 0 ~, \label{Kination Equation EFV}
\end{equation}
with primes denoting derivatives with respect to the number of e-folds of expansion, $dN_e=Hdt$. Since radiation becomes dominant at the end of the slow-roll regime, $H'/H = -2$, so that Eq.~\eqref{Kination Equation EFV} is $\phi''+\phi'=0$ and admits the simple solution:
\begin{equation}
\phi =  \phi'_0 \left(1-e^{-(N_e-N_0)}\right) ~, \label{Asymptotic Solution}
\end{equation}
where $N_0$ is the number of e-folds at which $\phi=0$. This implies that the field value quickly approaches $\phi_\infty= \phi'_0$, as anticipated above. We note that, more accurately $H'/H < -2$, since the inflaton's kinetic energy is not completely negligible, yielding slightly larger asymptotic field values, for instance $\phi_\infty\simeq 1.3 M_{\text{P}}$ in the example shown in Fig.~1. This nevertheless generically yields $\phi\sim M_P$ in this kinetic energy-dominated phase.

Since $H^2 \propto a^{-4}$ and $\phi' \propto \exp(-N_e) \propto a^{-1}$, the inflaton behaves as stiff matter ($w_{\phi}=1$), with $\rho_{\phi} \simeq H^2 \phi'^2/2 \propto a^{-6}$. The universe then undergoes a long radiation-dominated phase with a sub-leading inflaton stiff matter component (check $w_{\phi}$ in the bottom panel of Fig.~1). Eventually, the kinetic energy becomes diluted enough and the potential dominates the inflaton energy density after a number of e-folds since horizon-crossing $N_{\text{t}} \approx N_e + \log\left(K(0)/V(\phi_\infty)\right)/6$, where $K(0) = K(\phi=0)$ and $V(\phi_\infty) = \lambda m^{4+\alpha}/\phi_\infty^{\alpha}$. $N_{\text{t}}$, the total number of e-folds from horizon crossing, is $\sim100$ ($\log_{10}(1+\text{z}_{\text{t}}) \approx 10$), i.e.~shortly after BBN. From this point onwards, $V(\phi) \gg K(\phi)$, $w_{\phi} \approx -1$ and the inflaton fluid is indistinguishable from a cosmological constant because the Hubble friction term, $H$, is much larger than the effective mass of the field $m_{\phi}^2 = V''(\phi_\infty) =\alpha (\alpha + 1) V(\phi_\infty)/\phi_\infty^2 \sim \alpha(\alpha+1)H_0^2$. As a result, the field is effectively frozen, and only when $H \sim H_0$ can $\phi$ start to evolve significantly, as usual in thawing quintessence models\footnote{This occurs even though the inflaton potential is an inverse power law, usually associated with tracking freezing models of quintessence \cite{Tsujikawa:2013fta}. This shows the importance of initial conditions in quintessence models of dark energy. This behavior had been pointed out earlier in \cite{Peebles:1998qn}.}. This means that the cosmological evolution of this model is somewhat independent of the particular quintessence potential, as long as it decreases rapidly once the field crosses the origin. It could, therefore, be replicated in more well-motivated potentials \cite{Tashiro:2003qp} and our analysis is not restricted to the particular Peebles-Vilenkin form.

\begin{figure}
	\centering\includegraphics[scale=0.60]{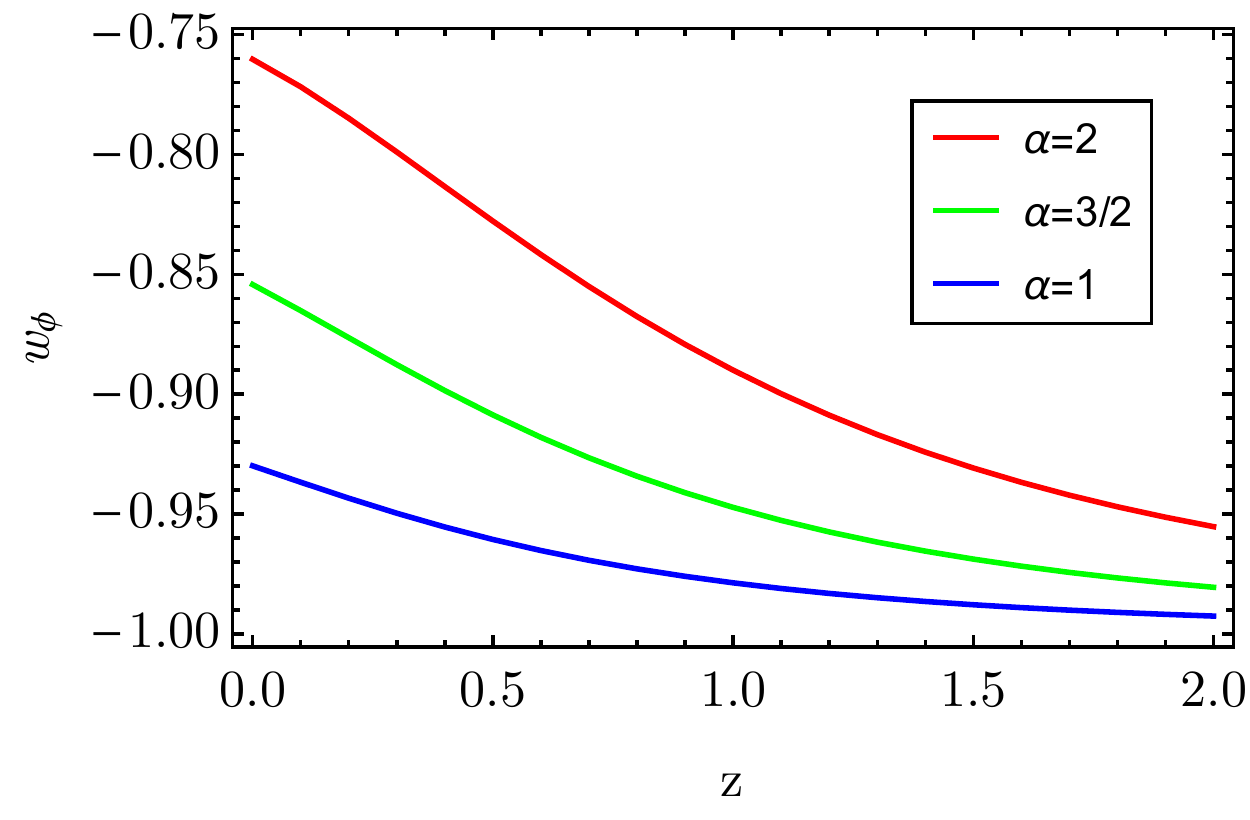} 
	\caption{Evolution of the quintessence equation of state at late times for different values of $\alpha$ and the parametric choices in Figs.~1 and 2.}
\end{figure}

Due to the dilution of radiation and matter, the inflaton field eventually becomes the dominant component, causing the present accelerated expansion. Since the slow-roll parameters $\epsilon_{\phi} = \alpha^2 M_{\text{P}}^2 /(2\phi^2)$ and $\eta_{\phi} = \alpha(\alpha + 1) M_{\text{P}}^2 /\phi^2$ are non-negligible when the field becomes dynamical, $m_{\phi} \sim H$, the inflaton equation of state deviates from $w_{\phi} \approx -1$ due to an increase of $K(\phi)$:
\begin{equation}
w_{\phi} = \frac{K(\phi)-V(\phi)}{K(\phi)+V(\phi)} \approx -1 + \frac{K(\phi)}{V(\phi)} ~,
\end{equation}
where we have expanded the denominator for $K(\phi) \ll V(\phi)$. It relaxes back to $w_{\phi} \approx -1$ as the field dominates the energy budget of the Universe, according to the de Sitter no-hair theorem \cite{Wald:1983ky}, as the field moves towards larger values where $\epsilon_\phi\ll 1$.

Choosing different values of $\alpha$ or $Q_*$ (and therefore $\phi'_0$) will result in different late time behaviors, as represented in Fig.~3. The $(w_0, w_a)$ parameterization \cite{Linder:2015zxa}, $w_{\phi} = w_0 + w_a(1-a)$ with $a = 1/(1-z)$, conveniently encapsulates the variation of $w_{\phi}$,  yielding in this example $(-0.73,-0.33)$, $(-0.85,-0.20)$, $(-0.93,-0.09)$ for $\alpha = 2,~3/2,~1$, respectively\footnote{It was mentioned in \cite{Linder:2015zxa} that a large range of thawing models is well fitted by $w_a \approx -1.58 (1 + w_0)$. We observe a slightly smaller slope, $w_a \approx -1.26 (1 + w_0)$, for the examples of Fig.~3.
}. As expected, smaller values of $\alpha$ correspond to shallower quintessence potentials (smaller $\epsilon_{\phi}$, $\eta_{\phi}$), which lead to smaller deviations from $w_{\phi} \approx -1$ at late times. This has some degeneracy with the dissipation at horizon crossing, $Q_*$, as larger values of this quantity lead to a larger $\phi'_0$ and consequently a value of $w_{\phi}$ closer to a cosmological constant. For instance, choosing $Q_* = 0.07$,  we find $\phi'_0 \approx 1.3 M_{\text{P}}$ and so $(w_0, w_a)$ becomes $(-0.77,-0.29)$, $(-0.87,-0.17)$ , $(-0.94,-0.08)$ for $\alpha = 2,~3/2,~1$. Comparing the above results with bounds on $(w_0, w_a)$ from the \textit{Planck} collaboration \cite{Ade:2015rim,Aghanim:2018eyx} suggests that $\alpha = 2$ ($\alpha = 3/2$) is (marginally) outside the $68\%$ confidence interval for the \textit{Planck}+SNe+BAO data, with $\alpha = 1$ inside it, but these conclusions should be taken at face value, given the particular details of the model and the data-set dependence of the $(w_0, w_a)$ bounds (see e.g.~\cite{DiValentino:2017zyq}).

In summary, in this Letter, we have shown that the inflaton can account for the present accelerated expansion of the Universe in the context of a concrete quantum field theory model, the Warm Little Inflaton scenario \cite{Bastero-Gil:2016qru}, as a result of the smooth transition between a period of warm inflation and a radiation-dominated era and the fact that the symmetries of the model make the inflaton both stable and sterile at late times. Alongside our earlier work on inflaton-dark matter \cite{Rosa:2018iff} and previous works on baryogenesis/mattergenesis during warm inflation \cite{BasteroGil:2011cx, Bastero-Gil:2014oga, Bastero-Gil:2017yzb}, this shows that  this alternative paradigm may have a very significant impact not only on the inflationary dynamics itself but also on the other major open problems in cosmology.

The dynamics provides several interesting observational features: warm inflation generically predicts a suppressed tensor-to-scalar ratio and a modified consistency relation between the latter and the tensor spectral index \cite{BasteroGil:2009ec, Cai:2010wt, Bartrum:2013fia}, which will be further probed with next generation CMB experiments \cite{Abazajian:2016yjj,Ade:2018sbj}. Furthermore, the current data already constrains the consistent $(w_0, w_a)$ space of models of quintessence severely, causing them to be very close to a cosmological constant (this depends somewhat on the analysis, see \cite{DiValentino:2017zyq,Tosone:2018qei}). This causes smaller values of $\alpha$ to be favored in the particular scenario displayed here. Note, however, that this conclusion is quite model-dependent, as scenarios which predict larger values of $Q_*$ will consequently have larger asymptotic field values, allowing for a larger range for the consistent values of $\alpha$.

 We also emphasize that our main results are not restricted to the particular Peebles-Vilenkin form of the quintessence part of the potential nor on how it is "glued" to the quartic inflationary potential, since the dynamics is insensitive to the potential function for $0<\phi\lesssim M_{\text{P}}$, and the exponent $\alpha$ simply characterizes the slope of the potential at $\phi\sim \phi_\infty \sim M_{\text{P}}$.
 
 Finally, a potentially distinctive feature of our scenario are dark energy isocurvature fluctuations, which would be (anti-)correlated with the main adiabatic curvature perturbations given their common origin in inflaton fluctuations. The thermal nature of such fluctuations, with a distinctive spectrum dependent on the form of the dissipation coefficient during inflation ($\Upsilon\propto T$ in this case), could also in principle be used to probe the warm vs.~cold nature of the quintessential inflation scenario. Detecting dark energy isocurvature modes may, however, be quite hard in general, given that they only affect the CMB spectrum on large scales through the integrated Sachs-Wolfe effect at late times (see e.g.~\cite{Kawasaki:2001nx, Moroi:2003pq, Matsui:2018xwa}), but we hope that the present work motivates further exploration of this possibility.

\section*{Acknowledgements}
L.B.V. is supported by FCT grant PD/BD/140917/2019. J.G.R. is supported by the FCT Investigator Grant No. IF/01597/ 2015. This work was partially supported by the H2020-MSCA-RISE-2015 Grant No. StronGrHEP-690904 and by the CIDMA Project
No. UID/MAT/04106/2019. L.B.V. thanks Tiago Barreiro and Juan P. Beltr\'an Almeida for private correspondence.

\vspace{1cm}

During the completion of this work, we became aware of the recent analysis of warm quintessential inflation in \cite{Dimopoulos:2019gpz}, which shares some similarities with this Letter, albeit using a different but nevertheless complementary approach. Here we consider a concrete quantum field theory model of inflation, which allows us to follow in detail the dynamical evolution of dissipative effects. This affects not only the duration of inflation and associated observational predictions but also the quintessence phase at late times. In particular, in our scenario the non-negligible dissipation at the end of inflation prevents an intermediate kination era, so that the field cannot evolve towards large super-Planckian values, thus requiring shallower quintessence potentials.



\end{document}